\documentclass[a4paper, 10pt, conference]{ieeeconf}      % Use this line for a4
\usepackage{comment}
\usepackage{array}
\usepackage{cite}
\usepackage{upgreek}
\usepackage{amsmath,amssymb,amsfonts}
\usepackage{algorithmic}
\usepackage{graphicx}
\usepackage{multirow}
\usepackage{textcomp}
\usepackage{xcolor}

\IEEEoverridecommandlockouts                              % This command is only
\makeatletter
\let\NAT@parse\undefined
\makeatother
\usepackage{hyperref}
\title{\LARGE \bf
Mood as a Contextual Cue for Improved Emotion Inference
}

\author{\parbox{16cm}{\centering
    {\large Soujanya Narayana, Ibrahim Radwan, Ramanathan Subramanian, Roland Goecke}\\
    {\normalsize
    University of Canberra, ACT, Australia\\}}
}

\begin{document}

\begin{comment}
\ifFGfinal
\thispagestyle{empty}
\pagestyle{empty}
\else
\author{Anonymous FG2024 submission\\ Paper ID \FGPaperID \\}
\pagestyle{plain}
\fi
\end{comment}
\maketitle

%%%%%%%%%%%%%%%%%%%%%%%%%%%%%%%%%%%%%%%%%%%%%%%%%%%%%%%%%%%%%%%%%%%%%%%%%%%%%%%%%%%%%%%%%%%%%%%%%%%%%%%%%%

\begin{abstract}
Psychological studies observe that emotions are rarely expressed in isolation and are typically influenced by the surrounding context. While recent studies effectively harness uni- and multimodal cues for emotion inference, hardly any study has considered the effect of long-term affect, or \emph{mood}, on short-term \emph{emotion} inference. This study (a) proposes time-continuous \emph{valence} prediction from videos, fusing multimodal cues including \emph{mood} and \emph{emotion-change} ($\Delta$) labels, (b) serially integrates spatial and channel attention for improved inference, and (c) demonstrates algorithmic generalisability with experiments on the \emph{EMMA} and \emph{AffWild2} datasets. Empirical results affirm that utilising mood labels is highly beneficial for dynamic valence prediction. Comparing \emph{unimodal} (training only with mood labels) vs \emph{multimodal} (training with mood and $\Delta$ labels) results, inference performance improves for the latter, conveying that both long and short-term contextual cues are critical for time-continuous emotion inference. 

\end{abstract}

%%%%%%%%%%%%%%%%%%%%%%%%%%%%%%%%%%%%%%%%%%%%%%%%%%%%%%%%%%%%%%%%%%%%%%%%%%%%%%%%%%%%%%%%%%%%%%%%%%%%%%%%%

\section{INTRODUCTION}
\label{sec::introduction}

%Context both produces emotion, and shapes how it is interpreted. 
Affective technologies have made significant strides over the past decade. Their primary aim is to recognise emotions employing multimodal cues, which enables effective human-computer interaction. Emotionally-aware systems are crucial to understanding human behaviour, and uninterrupted interaction between humans and computers~\cite{beale2008role}. 
%Numerous recent studies use multimodal cues like face, voice, biosignals, etc. to infer emotions~\cite{wei2020multimodal, bhattacharya2021exploring, pini2017modeling}. Since modalities can complement each other semantically, it is evident that multimodal systems are more robust than the unimodal counterparts~\cite{busso2004analysis, huang2020multimodal}. 
While behavioural modalities such as face, voice, biosignals, \emph{etc.} are beneficial for emotion inference~\cite{wei2020multimodal, bhattacharya2021exploring}, psychology studies corroborate that in myriad ways, \emph{context} influences cognition, emotion, behaviour, and their perception~\cite{mcnulty2012beyond, barrett2011context}. Psychological processes cannot be interpreted independently of the context in which they occur. \emph{Frege's Context Principle} states that one should not seek the meaning of the words in isolation, but in the context of the sentences in which they appear~\cite{dummett1993context}. Philosophical and psychological studies emphasise that context plays a central role in emotion research~\cite{greenaway2018context} and emotion regulation~\cite{aldao2013future}. 

From a computational viewpoint, recent studies have made considerable progress towards context-aware emotion inference. Inspired by the psychology literature, several studies have employed multimodal cues -- faces and pose/gait features~\cite{mittal2020emoticon}, scene context~\cite{kosti2017emotion, thuseethan2022emosec}, group-level features~\cite{guo2017group, abbas2017group}, inter-agent interactions~\cite{mittal2020emoticon}, \emph{etc.} -- to model contextual information in images and videos to infer emotions. Psychologists have comprehensively examined the interactions between mood and emotion. However, contemporary emotion inference methods have disregarded the contribution of \emph{mood} as a contextual cue while predicting emotions. Whilst the terms \emph{mood} and \emph{emotion} are often used synonymously~\cite{katsimerou2015predicting}, they are closely related but distinct affective phenomena. They differ in terms of duration, intensity and attributes. While emotion is an induced and short-term affective state, mood is regarded as a diffused and long-term affective state~\cite{jenkins1998human}. 

Besides influencing cognitive functions such as evaluative judgements and memory retrieval~\cite{picard2000affective}, mood is also known to impact human emotion recognition~\cite{siemer2001mood}. The authors of~\cite{wong2016mood} propose the mood-emotion loop, which states that mood is a higher-order variable that activates lower-level states such as emotions. Furthermore, they argue that these two affective mechanisms form a loop, \emph{i.e.}, repeatedly triggering each other. A mood-congruity effect is proposed in~\cite{schmid2010mood}, which states that mood creates a bias, hampering the recognition of an incongruent emotion.

\begin{figure*}[!ht]
\centering
\includegraphics[width=0.95\textwidth]{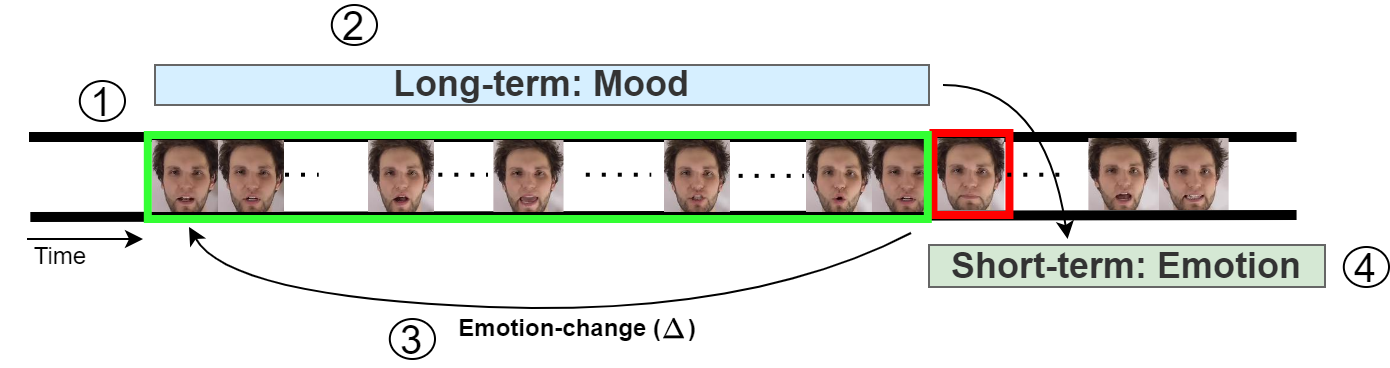}\vspace{-2mm}
\caption{\textbf{Problem Overview: We perform emotion inference in videos using mood and emotion change ($\Delta$) labels as contextual cues. (1) For each clip (an exemplar clip segment from the AffWild2 dataset~\cite{kollias2019expression} is denoted in green), we utilise existing mood labels or derive them from valence annotations. (2) Moreover, we utilise emotion-change ($\Delta$) information in the form of the valence differential between the first and last frames of the video clip. (3) Employing mood and emotion-change labels, we predict the valence rating for the succeeding frame (red).}}\vspace{-3mm}
\label{fig:overview}
\end{figure*}

Inspired by compelling evidence from psychology regarding the mood-emotion interplay, we investigate the utility of \emph{mood} as a \emph{contextual cue} for inferring emotional valence. Furthermore, we examine \emph{emotion-change} ($\Delta$) information as an additional cue. \emph{Valence} refers to the degree of pleasantness or unpleasantness following exposure to a stimulus or an event~\cite{russell1980circumplex}, while $\Delta$ denotes the difference in valence labels between frames~\cite{narayana2022improve, narayana2023weakly}. Fig.~\ref{fig:overview} overviews our study, where we examine the cumulative impact of long-term affect (mood) and changes in short-term affect ($\Delta$), to infer valence at the subsequent time step.     

In this work, we perform emotion inference on the \emph{EMMA}~\cite{katsimerou2016crowdsourcing} and \emph{AffWild2}~\cite{kollias2019expression} video datasets with mood and $\Delta$ labels as contextual cues. Mood labels are derived from valence annotations for the AffWild2 database, while ground-truth mood labels are utilised for EMMA. A two-branch 3-dimensional Convolutional Neural Network (3D-CNN) is trained with mood labels and valence labels, with 3D ResNet-18~\cite{hara2017learning} and EmoFAN~\cite{toisoul2021estimation} as encoders in each branch respectively, followed by a projection head. This is compared with a three-branch 3D-CNN trained with mood, $\Delta$, and valence labels, with ResNet-18 as encoder for the first two branches, and EmoFAN as encoder for the last branch, followed by a projection head. Finally, spatial, channel, and temporal attention modules are integrated independently as well as in combination within these models.

Empirical results confirm that emotion inference improves when both mood and emotion-change labels are employed as contextual cues. Additionally, integrating various attention modules further enhances emotion prediction performance. We summarise our key research contributions as follows:

\begin{enumerate}
    \item We propose to infer emotions by using mood and emotion-change labels as \emph{context}. To this end, we perform a regression task by training a two-branch 3D-CNN with mood and valence labels, and a three-branch 3D-CNN with mood, $\Delta$, and valence labels. 
    \item We integrate spatial, channel, and temporal attention modules independently and in combination for improved emotion inference.
    \item Experiments reveal that mood and $\Delta$ information as context improve emotion prediction. The positive impact of mood and $\Delta$ labels is noted even with the attention modules integrated.
    \item We observe identical trends on the diverse EMMA~\cite{katsimerou2016crowdsourcing} and AffWild2~\cite{kollias2019expression} datasets, implying that our proposed methods are generalisable.
    \item We verify the effectiveness of attention modules and their respective arrangements within the models via an ablative study.
\end{enumerate}

%%%%%%%%%%%%%%%%%%%%%%%%%%%%%%%%%%%%%%%%%%%%%%%%%%%%%%%%%%%%%%%%%%%%%%%%%%%%%%%%%%%%%%%%%%%%%%%%%%%%%%%%%%%%

\section{RELATED WORK}
\label{sec::related_work}

%Numerous studies have explored various modalities for inferring emotions by employing various machine learning approaches on diverse affective databases.
This section briefly overviews (a) automatic emotion estimation via various modalities (Sec.~\ref{subsec::automatic_emotion}), (b) emotion prediction employing various contexts (Sec.~\ref{subsec::context_emotion}), and (c) studies which examine the interaction between mood and emotion (Sec.~\ref{subsec::mood_emotion_interplay}). 

\subsection{Dimensional Emotion Estimation}
\label{subsec::automatic_emotion}

The \emph{dimensional} model of emotion posits that emotions can be represented along multiple continuous dimensions, as opposed to the \emph{categorical} model, which asserts the existence of seven basic, discrete and universally recognised emotions namely, \emph{happiness, sadness, anger, fear, disgust, and contempt} and \emph{surprise}~\cite{ekman1969pan}. While multiple dimensional models were proposed, the most widely accepted is the \emph{Circumplex model of affect}, which posits that emotions lie in a circular space spanned by two dimensions, \emph{valence} on the horizontal axis, denoting the degree to pleasantness or unpleasantness, and \emph{arousal} on the vertical axis, referring to the degree to excitation or calmness~\cite{russell1980circumplex} following exposure to a stimulus/event. Dimensional models overcome the restrictive nature of discrete emotions, as they capture subtle gradations offering a more precise and fine-grained representation of emotions, whereas categories oversimplify human affective states~\cite{russell1980circumplex}. Additionally, the range of emotional displays of humans on a daily basis cannot be manifested by coarse emotional classes, and is better represented as a continuous spectrum~\cite{posner2005circumplex}.

The success of deep learning approaches has led to their extensive use in facial affect analysis. The Emotion-Focussed Attention Network (EmoFAN)~\cite{toisoul2021estimation} employs a neural attention mechanism on top of a Face Alignment Network (FAN)~\cite{bulat2017far}, to jointly predict discrete and continuous emotions. A comparative analysis of the performance of Recurrent Neural Networks, self-attention mechanisms, and cross-modal attention for audio-visual affect recognition on AffWild2~\cite{kollias2019expression} is undertaken in~\cite{karas2022time}. In~\cite{choi2020multimodal}, a multimodal attention network, integrating facial video features and EEG signals, is employed for continuous emotion inference, and is observed to outperform its unimodal counterpart. Semi-supervised learning, a popular approach to address the annotation of huge affective datasets, is adopted in~\cite{choi2020semi} and~\cite{parameshwara2023efficient} for video-based emotion inference. Additionally in~\cite{parameshwara2023examining}, few-shot learning is proposed as an alternative to reduce the burden of annotating large-scale dynamic  facial affect databases.

%-----------------------------------------------------------------------------------------------------------

\subsection{Context Modelling for Emotion Inference}
\label{subsec::context_emotion}
Several psychologists have conceptualised 
affective context in diverse ways. Viewed in its entirety, the main sources of context are (a) effects \emph{within the subject expressing emotion}, which refer to intrinsic cues of the subject such as the tone of the voice, (b) effects \emph{external to the subject expressing emotion}, which refer to the environmental cues such as scene information, people, \emph{etc.}, and (c) effects \emph{within the subject perceiving the emotion}, which refer to the perceiver-driven aspects such as age, social bias, \emph{etc.}~\cite{aviezer2017inherently}. 

Multiple emotion inference studies have recognised the significance of incorporating context information. In~\cite{kosti2019context}, the authors provide the EMOTIC database with discrete and continuous emotion annotations, and consider the scene information as context to infer the subject's emotion. A similar approach is followed in~\cite{lee2019context}, where the face-encoding and context-encoding streams are respectively employed to capture the facial region and relevant scene context by hiding the human face. The authors also provide the CAER database annotated with the six universal emotion categories. In addition to using multiple modalities (face and gait), the background visual information and inter-agent interactions are used to model the semantic context in~\cite{mittal2020emoticon} to perform multimodal context-aware emotion inference. The authors further provide the GroupWalk database, comprising videos recorded in real-world settings and annotated with emotion categories. By considering global temporal dynamics as context, continuous emotions are inferred from videos in~\cite{tellamekala2022modelling}.    

%----------------------------------------------------------------------------------------------------------

\subsection{Emotion, Mood, and their Interplay}
\label{subsec::mood_emotion_interplay}
The relation between mood and emotion is a historical psychological problem~\cite{russell2003core}. Mood is considered to be a diffused and enduring affective state as opposed to emotion, whose duration is relatively short. An apparent cause triggering a certain mood might not be known, but stimuli/events tend to elicit emotions. Mood is generally of low intensity, while emotions are relatively intense affective states~\cite{scherer2005emotions}. A mood-congruity effect on emotion recognition is observed in~\cite{schmid2010mood}, where positive mood interferes with the recognition of mood-incongruent negative emotion (and negative mood impedes the recognition of mood-incongruent positive emotion). The mood-emotion loop theory proposes that mood and emotion as distinct affective mechanisms, repeatedly triggering the arousal of each other~\cite{wong2016mood}. 

Mood disorders, such as unipolar and bipolar depression, cause impairments in emotion processing abilities~\cite{panchal2019cognitive}. Through an eye-tracking study, the authors in~\cite{schmid2011mood} note that positive mood fosters a global information processing style, where information is processed holistically rather than focusing on specific details, as compared to a sad mood, during facial emotion recognition. Contemporary studies have employed supervised learning and examined that emotion-change information contributes positively to mood inference~\cite{narayana2022improve}. Furthermore, mood prediction performance improves when pseudo emotion-change labels generated using weak-supervision are incorporated in the model~\cite{narayana2023weakly}.  

%----------------------------------------------------------------------------------------------------------

\subsection{Novelty of our Study}\label{subsec::novelty}
A comprehensive study of the literature reveals that: (a) While numerous studies infer continuous emotions employing unimodal and multimodal cues, very few studies examine context-aware emotion inference; (b) A majority of computational studies incorporating context perform categorical rather than dimensional emotion prediction; and (c) While the mood-emotion interplay has been extensively studied in psychology, hardly any computational studies employ mood as a contextual cue for emotion inference. 

Differently, this study explores the utility of video-level \emph{mood labels} as \emph{context} to infer dynamic valence levels. This study proposes to (1) infer valence values using mood and $\Delta$ labels via multimodal fusion; (2) integrate multimodal attention modules for enhanced emotion inference; and (3) examine the generalisability of context-based emotion inference on the EMMA and AffWild2 datasets. Additionally, the impacts of the components of our inference framework are empirically examined through an ablative study.

%%%%%%%%%%%%%%%%%%%%%%%%%%%%%%%%%%%%%%%%%%%%%%%%%%%%%%%%%%%%%%%%%%%%%%%%%%%%%%%%%%%%%%%%%%%%%%%%%%%%%%%%%%%

\section{MATERIALS}
\label{sec::materials}

This section details the datasets used (Section~\ref{subsec::datasets}), the labels used for model training (Section~\ref{subsec::labels}), and the procedure employed for generating input samples (Section~\ref{subsec::input_samples}). 

\subsection{Datasets}
\label{subsec::datasets}

\begin{figure*}[!ht]
\centering
\includegraphics[width=0.95\textwidth]{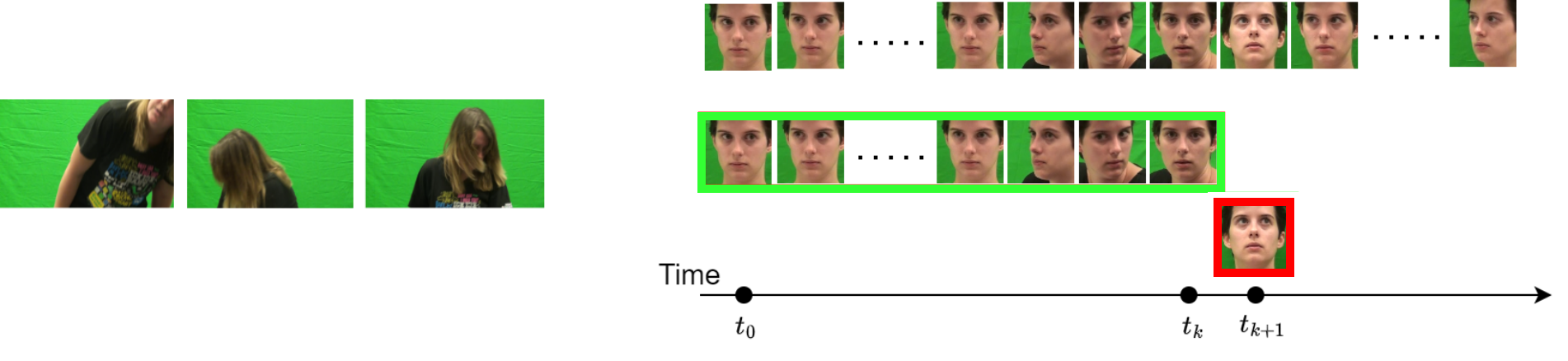}\vspace{-3mm}
\caption{\textbf{(Left) Exemplar frames in EMMA~\cite{katsimerou2016crowdsourcing}, showing an occluded/non-frontal face. (Right) Illustration of an input sample where the raw video is face-cropped, aligned and sub-sampled (top) to generate the visual information sequence from $t_0$ to $t_k$ for contextual emotion inference (middle), based on which valence at $t_{k+1}$ is inferred (bottom).}}
\label{fig:input_sample}\vspace{-2mm}
\end{figure*}

\subsubsection{\textbf{EMMA}} The  EMMA~\cite{katsimerou2016crowdsourcing} affective video dataset is a collection of long videos depicting improvised acting of non-interactive daily routines. 15 Dutch actors (6 male, 9 female) were recruited and were provided with the situational context and scenarios, followed by the acting-based mood induction procedure. These videos are labelled by human annotators via crowd-sourcing platforms such as \emph{Microworkers} for mood and emotion. Each video has a mood class annotation, namely, \emph{positive}, \emph{negative}, or \emph{neutral}, and continuous valence and arousal annotations in the range $[-1, 1]$. The dataset comprises 180 videos whose durations range from 0.33--5.23 minutes, with a mean of 2.03 minutes. The dataset is split into training, validation, and test sets in a subject-independent manner, to ensure that a subject appears in only one of these splits.     

\subsubsection{\textbf{AffWild2}} AffWild2~\cite{kollias2019expression} is a publicly available \emph{in-the-wild} affective video database capturing facial expressions. It comprises 558 videos collected from YouTube with 2,786,201 frames depicting human behaviour in real-world scenarios, with a total of 458 subjects (279 male, 179 female). The duration of these videos range from 0.03--26.22 minutes, with an average of 13.12 minutes. 

Four experts (2 male, 2 female) have annotated AffWild2 with continuous valence and arousal values in the range $[-1, 1]$, and the mean value of the four annotations is taken to be the final rating. Frames with annotations outside $[-1, 1]$ are not considered in the study. Additionally, all videos are annotated with emotion categories \emph{\emph{i.e.}, happiness, sadness, disgust, fear, anger, surprise, neutral, and other} by three experts. Akin to EMMA, AffWild2 is also split into training, validation, and test sets in a subject-independent manner. The  train, validation, and test sets comprise 341, 70, and 138 videos, respectively. Since the test set is not made publicly available, we use the validation set for evaluating our approach. 

%---------------------------------------------------------------------------------------------------------
\subsection{Labels}\label{subsec::labels}

\subsubsection{\textbf{Mood Labels}}
Since videos in the EMMA dataset are annotated with mood labels, we use them as ground-truth labels for model training. Each video has a mood label corresponding to the \emph{positive} $(+1)$, \emph{negative} $(-1)$, or \emph{neutral} $(0)$ categories. 

For AffWild2, we follow the approach outlined in~\cite{narayana2022improve} to derive mood labels. We derive mood annotations from frame-level valence ratings. Each video is assigned a mood label, which is either \emph{positive} $(+1)$, \emph{negative} $(-1)$, or \emph{neutral} $(0)$, from the valence ratings. Mood is considered to be a long-term affect~\cite{picard2000affective} and, aligned with this notion, the valence annotation persisting over the highest number of consecutive frames (or maximum duration) is utilised to assign the mood label. The mood label is assigned to $+1$, $-1$, or $0$ if the valence value persisting over most consecutive frames lies in the range $(0.3, 1]$, $[-0.3, -1)$, and $[-0.3, 0.3]$, respectively. 

\subsubsection{\textbf{Emotion-change ($\Delta$) Labels}}
In addition to using mood as a context, we also consider changes in emotion from time $t_0$ to $t_k$ to infer the valence at $t_{k+1}$. Since the valence values lie in the range $[-1, 1]$, the valence differential values are in the range $[-2, 2]$. Following~\cite{narayana2022improve}, the sign of the valence differential is assigned as the $\Delta$ label, which is either \emph{positive} $(+1)$, \emph{negative} $(-1)$, or \emph{neutral} $(0)$. Both EMMA and AffWild2 are annotated with frame-level valence values, and an identical procedure is followed for obtaining $\Delta$ labels for both datasets.

\subsubsection{\textbf{Valence Labels}}
Frame-level valence annotations in EMMA and AffWild2 are utilised as ground-truth for the training and validation sets. 

%-----------------------------------------------------------------------------------------------------------

\subsection{Generating Input Samples for Emotion Inference}
\label{subsec::input_samples}

\subsubsection{\textbf{Face Extraction in EMMA}} The EMMA video recordings capture the upper body of the subjects, while the AffWild2 dataset provides the cropped and aligned faces for each video. To ensure data consistency across datasets, we extract faces from EMMA videos employing OpenFace~\cite{Openface}, an open-source toolkit for facial behaviour analysis. OpenFace performs face detection, detects facial landmarks, and aligns and extracts faces from video frames. A confidence score of $<0.85$ reflects those frames where candidate detections may not correspond to an actual face due to multiple reasons such as body motion, occlusions, non-frontal faces, \emph{etc}. Fig.~\ref{fig:input_sample} (left) illustrates noisy face detection in EMMA. To mitigate noise propagation during model training, we generate input visual sequences by omitting these frames.   

\subsubsection{\textbf{Input Sample Generation}} From the raw EMMA and AffWild2 videos, we generate clips as a sequence of sub-sampled frames. For a video $v$ with $N$ frames, we extract clips of incrementally increasing length with a stride $s$. As shown in Fig.~\ref{fig:input_sample} (right), to predict valence of the frame $t_{k+1}$, we use the clip from $t_0$ to $t_{k}$; to predict valence for frame $t_{k+s}$, we employ visual information from $t_0$ to $t_{k+(s-1)}$, and this process continues until the terminal frame $t_N$, for which information until $t_{N-1}$ is used. The clip frames are sub-sampled to maintain a balance between computational cost and maximising facial affect information. Regardless of the clip length, we sample $n$ equally-spaced frames from each clip. Each generated clip comprises frames of the parent video alone and no frames from other videos.  

Each generated clip $c$ is assigned the mood label for its parent video. Additionally, $c$ is assigned a $\Delta$ label, computed as the valence differential between the first and last frames of $c$ (see Sec.~\ref{subsec::labels}). Therefore, $c$ has a mood label $\in$ \{-1, 0, 1\}, and a $\Delta$ label $\in$ \{-1, 0, 1\}.  

%%%%%%%%%%%%%%%%%%%%%%%%%%%%%%%%%%%%%%%%%%%%%%%%%%%%%%%%%%%%%%%%%%%%%%%%%%%%%%%%%%%%%%%%%%%%%%%%%%%%%%%%%%%%%%%

\section{EMOTION-INFERENCE APPROACH}
\label{sec::emotion_inference}

\begin{figure*}[!ht]
\centering
\includegraphics[width=0.75\textwidth]{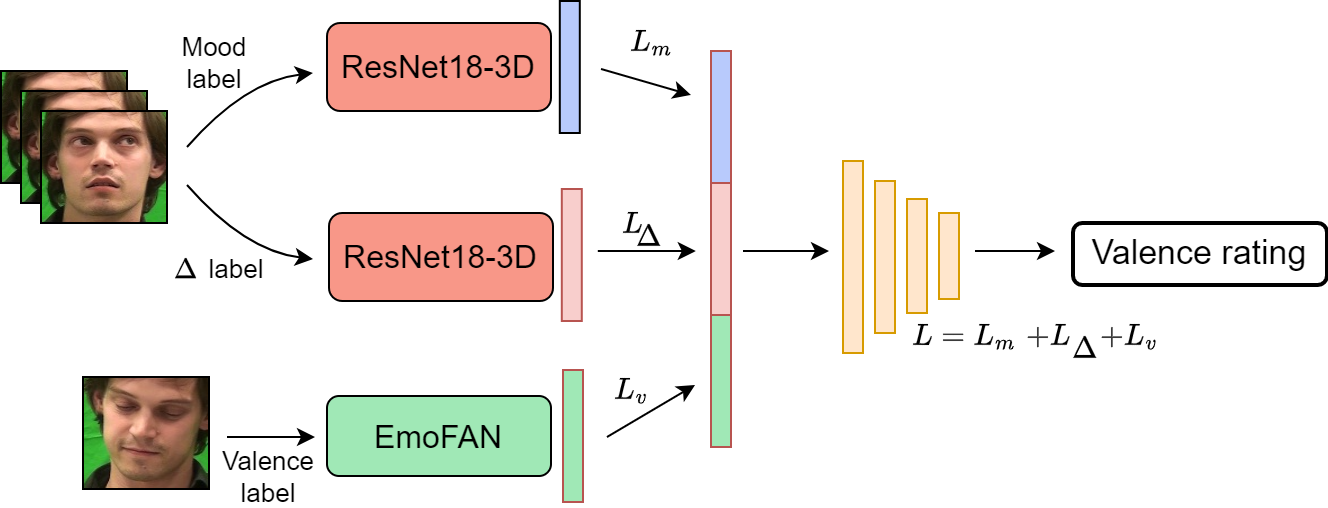}\vspace{-2mm}
\caption{\textbf{Architecture of the M$\Delta$-ValNet, a three-branch network utilising both mood and $\Delta$ labels for emotion inference. The architecture of ValNet comprises  only the bottom branch. Similarly, M-ValNet comprises the top and bottom branches depicted above.}}
\label{fig:mood_delta_valnet}\vspace{-2mm}
\end{figure*}

In a typical deep neural network trained end-to-end, a backbone is used to transform the raw input to meaningful representations, which are subsequently utilised for downstream tasks. In this section, we describe the various models used for emotion inference employing mood (and emotion change or $\Delta$) labels as context. 

%ResNet-3D (R3D), a 3-dimensional variant of the original ResNet architecture, is commonly employed as a backbone in emotion inference models~\cite{kuhnke2020two, jegorova2023ss}. EmoFAN~\cite{toisoul2021estimation}, built on top of the face alignment network for emotion inference, is another pretrained backbone model employed in multiple studies inferring affect~\cite{parameshwara2023efficient, sanchez2021affective}. 

\subsection{ValNet}\label{subsec::valnet}
To examine if mood as context is beneficial, we evaluate valence recognition performance with the \emph{ValNet} baseline, trained only with valence labels. Cropped and aligned faces extracted via the protocol employed in Sec.~\ref{subsec::input_samples}, denoted by red rectangles in Figures~\ref{fig:overview} and~\ref{fig:input_sample}, are input to this model. We employ EmoFAN~\cite{toisoul2021estimation}, built on top of a face alignment network, as the encoding backbone $\mathcal{B_V}$ as it demonstrates superior performance for emotion inference~\cite{parameshwara2023efficient, sanchez2021affective}. ValNet employs the pre-trained EmoFAN, $E(\cdot)$ as a backbone to map the input frame $x$ to a representation vector, $v = E(x) \in \mathcal{R^{D_{B_V}}}$, where $\mathcal{D_{B_V}} = 256$. This is followed by a projection head $P(\cdot)$,  which further maps $v$ to a vector $u = P(v) \in \mathcal{R^{D_P}}$, where $\mathcal{D_P} = 1$. $P(\cdot)$ is implemented via a Multi-Layer Perceptron (MLP), comprising four fully-connected (fc) layers with 256, 128, 128, and 1 neuron, respectively (denoted via cream-coloured rectangles in Fig.~\ref{fig:mood_delta_valnet}). The terminal fc layer neuron outputs the \emph{valence} rating. The input to each fc layer is batch normalised to achieve zero mean and unit variance, followed by ReLU activation. 

%-------------------------------------------------------------------------------------------------------------

\subsection{M-ValNet}\label{subsec::mood_valnet}
Utilising mood labels as context, we train the two-branch \emph{M-ValNet} composed of the top and bottom branches in Fig.~\ref{fig:mood_delta_valnet}. The first branch is fed with generated clips with mood labels denoted in green in Figures~\ref{fig:overview} and~\ref{fig:input_sample} (referred to as \emph{mood branch}), while the second branch is only fed with individual video frames along with their valence labels (henceforth referred to as the \emph{valence branch}). The mood branch employs Resnet-3D as the backbone (top branch of Fig.~\ref{fig:mood_delta_valnet}), denoted as $R3D(\cdot)$, to map the input clip $c$ to a vector, $u = R3D(c) \in \mathcal{R^{D_{B_M}}}$, where $\mathcal{D_{B_M}} = 256$. The valence branch comprises the \emph{ValNet} described above. Model-level fusion is  frequently used in affect inference, leveraging complementary information from multiple modalities to obtain a superior feature representation~\cite{praveen2021cross}. We fuse the mood and valence-specific representations by concatenating the vectors $u$ and $v$ to obtain $w = u \| v \in \mathcal{R^{D_F}}$, where $\mathcal{D_F} = 512$. A projection head $P(\cdot)$ further maps $w$ to a vector $z = P(w) \in \mathcal{R^{D_P}}$, where $\mathcal{D_P} = 4$. Similar to ValNet, $P(\cdot)$ is an MLP with four fc layers, respectively, comprising 256, 128, 128, and 4 neurons. Three neurons in the last fc layer correspond to the three mood classes, while the last neuron outputs the valence value. The input to each layer is batch normalised followed by ReLU activation.

%-------------------------------------------------------------------------------------------------------------

\subsection{M$\Delta$-ValNet}\label{subsec::mood_delta_valnet}
Besides employing mood as a context, we also utilise $\Delta$ labels for valence estimation via the \emph{M$\Delta$-ValNet} network.  \emph{M$\Delta$-ValNet} is a three-branch model as shown in Fig.~\ref{fig:mood_delta_valnet} fed with mood, $\Delta$, and valence labels, respectively. The mood and $\Delta$ branches take clips as input, while the valence branch is fed with individual frames and associated valence labels. The mood and $\Delta$ branches have identical ResNet-3D backbones $\mathcal{B_M}$ and $\mathcal{B}_{\Delta}$, respectively, followed by projection heads $\mathcal{P_M}(\cdot)$ and $\mathcal{P}_{\Delta}(\cdot)$.  Their combinations map the input clip $c$ to vectors $u = R3D(c) \in \mathcal{R^{D_{B_M}}}$ and $v = R3D(c) \in \mathcal{R^{D_{B}}}_{\Delta}$, where $\mathcal{D_{B_M}} = \mathcal{D_B}_{\Delta} = 256$. The valence branch uses EmoFAN as the backbone, $E(\cdot)$, to map the input frame $x$ to a vector $w = E(x) \in \mathcal{R^{D_{B_E}}}$, where $D_{B_E} = 256$. The three vectors $u$, $v$, and $w$ are concatenated to obtain a vector $y = u \| v \| w \in \mathcal{R^{D_F}}$, where $\mathcal{D_F} = 768$. A projection head $P(\cdot)$ further maps it to a vector $z = P(y) \in \mathcal{R^{D_P}}$, where $\mathcal{D_P} = 7$. $P(\cdot)$ is an MLP with four fc layers comprising 256, 128, 128, and 7 neurons, respectively. Three neurons in the last fc layer correspond to the mood classes, while the following three neurons correspond to the three $\Delta$ classes, and the last neuron outputs valence. The input to each layer is batch normalised followed by ReLU activation. 

%-------------------------------------------------------------------------------------------------------------

\subsection{Attention Modules}
\label{subsec::attention_modules}

\begin{figure*}[!ht]
\centering
\includegraphics[width=0.68\textwidth]{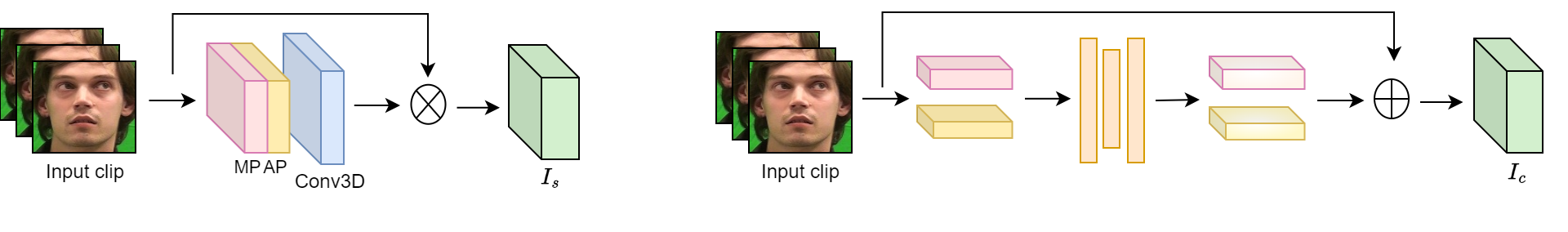}\vspace{-2mm}
\caption{\textbf{(Left) Spatial attention module. (Right) Channel attention module.}}
\label{fig:attention_modules}
\vspace{2mm}
% \end{figure*}

% \begin{figure*}[!ht]
\centering
\includegraphics[width=0.8\textwidth]{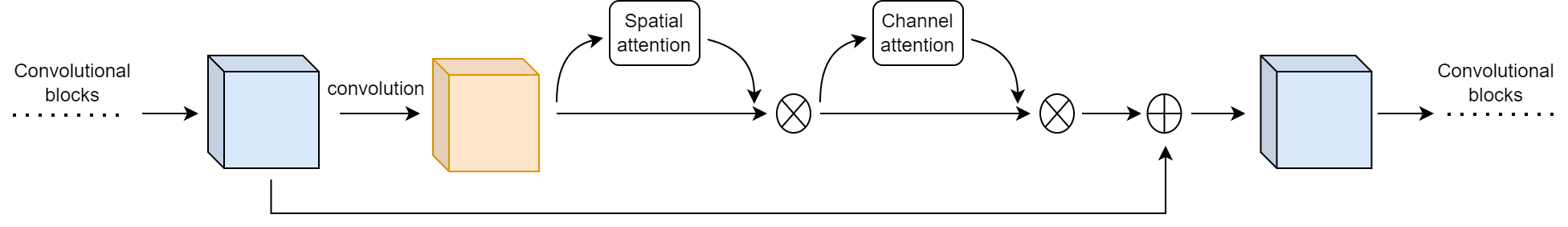}\vspace{-2mm}
\caption{\textbf{Illustration of the sequential arrangement of spatial and channel attention modules within a ResNet block.}}
\label{fig:resnet_cbam}
\end{figure*}

Similar to the human cognitive process of \emph{attention}, the neural attention mechanism enables a model to selectively focus on certain parts of the input data via importance weights. The Convolutional Block Attention Module (CBAM)~\cite{woo2018cbam} incorporates both spatial and channel-wise attention mechanisms within a CNN. Inspired by CBAM, we integrate spatial and channel attention modules into our models. 

\subsubsection{\textbf{Spatial Attention}}
Spatial attention enables to focus on relevant spatial locations or regions within the input data. CBAM consists of a spatial attention submodule, as shown in Fig.~\ref{fig:attention_modules} (left). The inter-spatial feature relationships are used to generate spatial attention maps. To compute spatial attention, average-pooling (AP) and max-pooling (MP) operations are applied along the channel axis. The obtained features are concatenated to generate an efficient feature descriptor, followed by a convolution operation on the concatenated vector to produce the spatial attention map. 

\subsubsection{\textbf{Channel Attention}} 
Channel attention allows the network to selectively emphasise or suppress certain channels within the feature maps. The channel attention module employed in CBAM, shown in Fig.~\ref{fig:attention_modules} (right), squeezes the spatial dimension of the input feature by implementing an average and max-pooling operations simultaneously. The resulting descriptors are further fed to a shared network comprising an MLP with one hidden layer. Element-wise summation is performed on the output vectors of the shared network to produce the channel attention map. 

\subsubsection{\textbf{Arrangement}}
Attention modules can either be stacked \emph{sequentially} one after the other, or can be arranged \emph{parallelly} as branches. As sequential arrangements have shown superior performance~\cite{woo2018cbam}, we follow a sequential arrangement of the attention modules, with spatial followed by channel attention. 

%------------------------------------------------------------------------------------------------------------------

\subsection{Performance Metrics}
Since we target a \emph{regression} task, we employ the \emph{Concordance Correlation Coefficient (CCC)} metric for performance evaluation as in~\cite{praveen2021cross, parameshwara2023efficient}. CCC inherently incorporates the Pearson Correlation Coefficient (PCC) and assesses the degree of agreement between two sets of measurements, evaluating how well they align along the $45^{\circ}$ line. If $y$ and $\hat{y}$ denote the ground truth and the predicted labels, respectively, PCC and CCC are defined as, 
\label{eq:cccloss}
\begin{equation}
PCC(y, \hat{y}) = \frac{\mathcal{E}[(y - \mu_{y})(\hat{y} - \mu_{\hat{y}})]}{\sigma_{y}\sigma_{\hat{y}}} 
\end{equation}
\begin{equation}
CCC(y, \hat{y}) = \frac{2\sigma_{y}\sigma_{\hat{y}}PCC(y, \hat{y})}{\sigma_{y}^2 + \sigma_{\hat{y}}^2 + (\mu_{y} - \mu_{\hat{y}})^2},
\end{equation}

\noindent where $\mu_{y}$ and $\sigma_{y}$, respectively, denote the mean and the standard deviation of $y$, and $\mathcal{E}$ denotes the expected value.

%-------------------------------------------------------------------------------------------------------------------

\subsection{Loss Functions}\label{subsec::loss_functions}
For each model described in Sections~\ref{subsec::valnet}--\ref{subsec::mood_delta_valnet}, we employ a dynamically weighted loss function, as proposed in~\cite{parameshwara2023efficient}. \emph{E.g.}, in the valence branch, $\mathcal{L}_v$ is defined as,  
\begin{equation} 
    \mathcal{L}_v = f * \mathcal{L}_{MSE} + g * \mathcal{L}_{CCC}
\end{equation}

\noindent where $\mathcal{L}_{MSE}$ is the MSE loss, $\mathcal{L}_{CCC}$ = 1 - $CCC$, and $f$ and $g$ are dynamic weight functions given by $f = \alpha \left(\frac{i}{m}\right)^k$ and $g = 1 - \left(\frac{i}{m}\right)^k$, respectively, where $i$ denotes the $i^{th}$ training epoch from a total $m$ epochs, while $\alpha \in \mathcal{R}$ and $k \in \mathcal{Z^+}$ are hyper-parameters, respectively controlling normalisation and non-linearity.

Since the mood and $\Delta$ branches are designed to perform a classification task, we optimise the cross-entropy loss. In \emph{M-ValNet}, the losses for the mood ($\mathcal{L}_m$) and valence ($\mathcal{L}_v$) branches are summed up, and the cumulative loss, $\mathcal{L} = \mathcal{L}_m + \mathcal{L}_v$ is optimised. Similarly, in \textit{M$\Delta$-ValNet}, the loss from the three branches, $\mathcal{L}_m, \mathcal{L}_\Delta$ and $\mathcal{L}_v$ are summed up and the cumulative loss, $\mathcal{L} = \mathcal{L}_m +  \mathcal{L}_{\Delta} + \mathcal{L}_v$ is optimised (see Figure~\ref{fig:mood_delta_valnet}). 

%-----------------------------------------------------------------------------------------------------------------

\subsection{Implementation Details}
\label{subsec::implementation}

All experiments in this study are performed using the \emph{PyTorch} library (version 1.12). The models are trained on four NVIDIA GeForce RTX 2080 Ti with 12GB memory each. To generate the EMMA and AffWild2 input samples, we employ an initial temporal length (video duration) of 200 frames. Subsequent clips are generated by adding $n = 5$ sampled frames at a time with a stride $s = 3$. The models are trained using the Adam optimiser, with the learning rate reduced by a factor of 10 for every 10 epochs, and the base learning rate set to 0.001. The models are trained for 45 epochs with a batch size of 210, and the dropout rate set to 0.5.

%%%%%%%%%%%%%%%%%%%%%%%%%%%%%%%%%%%%%%%%%%%%%%%%%%%%%%%%%%%%%%%%%%%%%%%%%%%%%%%%%%%%%%%%%%%%%%%%%%%%%%%%%%%%%%%%%%

\section{RESULTS AND DISCUSSION}
\label{sec::results}

{\renewcommand{\arraystretch}{1.3}%
\begin{table*}[t]
\caption{Performance results of the various models on EMMA and AffWild2. Best results obtained without and with attention are highlighted using $^\dagger$ symbol and bold font, respectively.}
\centering
\label{tab:main_emma_affwild2}
\resizebox{0.82\textwidth}{!}{%
\begin{tabular}{|c|c|c|c|c|}
\hline 
\textbf{Model}               & \textbf{Training labels} & \textbf{Attention module} & \textbf{CCC (EMMA)} & \textbf{CCC (AffWild2)} \\ \hline \hline
ValNet               & Valence                 & -               & 0.33 & 0.23 \\
M-ValNet             & Mood, Valence           & -               & 0.36 & 0.43 \\
M$\Delta$-ValNet     & Mood, $\Delta$, Valence & -               & $0.53^\dagger$ & $0.44^\dagger$ \\
M-ValNet + Attention & Mood, Valence           & Spatial-Channel & 0.39 & 0.41 \\
M$\Delta$-ValNet + Attention & Mood, $\Delta$, Valence  & Spatial-Channel           & \textbf{0.57}                & \textbf{0.49}                    \\ \hline
\end{tabular}}
\end{table*}

\begin{table*}[t]
\caption{Comparison of different methods on the validation set of AffWild2. Best results are in bold, and second-best are underlined.}
\centering
\label{tab:comparison_sota}
\resizebox{0.82\textwidth}{!}{%
\begin{tabular}{|c|c|c|c|}
\hline
\textbf{Model}                             & \textbf{Context}     & \textbf{Modality}           & \textbf{CCC} \\ \hline \hline
Kollias \emph{et al.}~\cite{kollias2020analysing}                                   & -                    & Unimodal (Visual)           & 0.14                \\
Deng \emph{et al.}~\cite{deng2020multitask}        & -                    & Unimodal (Visual)           & 0.24                \\
Sanchez \emph{et al.}~\cite{sanchez2021affective}    & Temporal             & Unimodal (Visual)           & 0.44                \\
Tellamekala \emph{et al.}~\cite{tellamekala2022modelling} & Temporal             & Unimodal (Visual)           & \textbf{0.56}                \\ \hline
Zhang \emph{et al.}~\cite{zhang2020m}       & -                    & Multimodal (Audio + Visual) & 0.40                 \\
Kuhnke \emph{et al.}~\cite{kuhnke2020two}      & -                    & Multimodal (Audio + Visual) & \underline{0.49}                \\
Meng \emph{et al.}~\cite{meng2022multimodal}        & Temporal             & Multimodal (Audio + Visual) & \textbf{0.61}                \\ \hline
\textbf{M$\Delta$-ValNet + Attention (Ours)}        & Mood, Emotion-change & Unimodal (Visual)           & \underline{0.49}                \\ \hline
\end{tabular}}
\end{table*}
}

{\renewcommand{\arraystretch}{1.2}%
\begin{table*}[t]
\caption{Ablation study results conveying the impact of spatial-channel attention module placements.}
\centering
\label{tab:ablation_placement}
\resizebox{0.78\textwidth}{!}{%
\begin{tabular}{|c|c|c|c|c|}
\hline
\textbf{Model} &
  \textbf{Training labels} &
  \textbf{\begin{tabular}[c]{@{}c@{}}Placement of \\ attention modules\end{tabular}} &
  \textbf{CCC (EMMA)} &
  \textbf{CCC (AffWild2)} \\ \hline \hline
M-ValNet         & Mood, Valence           & Within ResNet  & 0.36 & 0.43 \\
                 &                         & Outside ResNet & 0.33 & 0.35 \\ \hline
M$\Delta$-ValNet & Mood, $\Delta$, Valence & Within ResNet  & 0.57 & 0.49 \\
                 &                         & Outside ResNet & 0.55 & 0.42 \\ \hline
\end{tabular}}\vspace{-2mm}
\end{table*}
}

Table~\ref{tab:main_emma_affwild2} presents CCC values obtained on the EMMA and AffWild2 datasets using various models, including and excluding attention modules. ValNet is a model trained without any context, M-ValNet is trained with both mood (as context) and valence labels, while M$\Delta$-ValNet is trained with mood, $\Delta$ labels (providing contextual information), plus valence labels. 

Considering models trained on the EMMA dataset, excluding the attention modules, a higher CCC is obtained with M-ValNet than ValNet. Furthermore, a significantly higher CCC is obtained with M$\Delta$-ValNet than both ValNet and M$\Delta$-ValNet, conveying that utilizing mood plus $\Delta$ labels as context is highly beneficial. When sequential spatial-channel attention modules are integrated with the models, a similar trend is observed, where a higher CCC is achieved with M$\Delta$-ValNet as compared to M-ValNet or ValNet. Identical trends are observed for the AffWild2 dataset, where models trained with mood and $\Delta$ labels result in a higher CCC, as compared to their counterparts trained only with mood labels. 

The observed results indicate that model training with additional contextual information improves emotion inference performance. Firstly, M-ValNet is fed clips with the associated mood labels as input, enabling the model to learn and utilize the affect prevalent over a longer period of time for valence inference.  Secondly, providing emotion-change information in addition to the mood label is akin to providing both, the integral and differential affective context. This enables the model to observe the dominant affect, along with the intricate changes in affect over time. Cumulatively, obtained results indicate the utility of mood and emotion-change information as context for valence prediction.

The efficacy of mood and emotion-change as context is also apparent when attention modules are integrated into the models. Results reveal that even upon incorporating attention modules, employing both mood and $\Delta$ labels as context improves emotion inference performance, as compared to employing only mood labels. Higher CCC values with the integration of attention can be attributed to the nature of spatial and channel attention, which  respectively enable selective focus on relevant regions and the most informative input channels.

Highly similar trends noted for the AffWild2 dataset demonstrate generalisability of our emotion-inference approach. While the EMMA dataset is collected in a controlled setting, AffWild2 represents an \textit{in-the-wild} database. It is noteworthy that our approach of using both mood and emotion-change information as context for emotion inference shows consistent trends across diverse datasets. Overall, our results confirm that modelling mood and emotion-change as context contribute positively to emotion inference, with or without attention, and that our valence inference framework is generalisable.    

Table~\ref{tab:comparison_sota} compares our results with the state-of-the-art (SOTA) on the AffWild2 validation set. The first four rows indicate the results of unimodal methods considering visual cues, and the following three rows indicate the results obtained with multimodal methods fusing audio and visual cues. Firstly, our proposed approach outperforms the baseline result of~\cite{kollias2020analysing} and the performance reported in~\cite{deng2020multitask}, where emotion is inferred without any contextual information. Using mood and emotion-change information as context results in better emotion inference as compared to~\cite{sanchez2021affective}, where temporal context is modelled. Comparing with the unimodal methods, our approach performs comparable to SOTA. 

Our method also yields the second-best CCC along with~\cite{kuhnke2020two} among multimodal approaches. We obtain a much higher CCC than~\cite{zhang2020m}, employing audio-visual cues for valence estimation. Overall, our notion of considering mood and emotion-change labels as contextual cue achieves performance very comparable to uni and multimodal methods ignoring context. Even among studies considering temporal context (\cite{tellamekala2022modelling},~\cite{sanchez2021affective},~\cite{meng2022multimodal}), our approach yields comparable performance (see rows 3, 4, and 7 of Table~\ref{tab:comparison_sota}). 

\subsection{Ablation Study}
%The findings reported in Table~\ref{tab:main_emma_affwild2} are corroborated through an ablation study conducted on EMMA and the efficacy of the various components of the proposed approach are examined.  
We performed an ablation study on EMMA to corroborate the findings shown in Table~\ref{tab:main_emma_affwild2}, and to examine the efficacy of the various architectural components.

{\renewcommand{\arraystretch}{1.25}%
\begin{table*}[t]
\caption{Ablation study results using different attention modules.}
\centering
\label{tab:ablation_attention_modules}
\resizebox{0.65\textwidth}{!}{%
\begin{tabular}{|c|c|c|c|}
\hline
\textbf{Model}   & \textbf{Training labels} & \textbf{Attention} & \textbf{CCC (EMMA)} \\ \hline \hline
M-ValNet         & Mood, Valence            & Spatial            & 0.34                \\
                 &                          & Temporal           & 0.35                \\
                 &                          & Channel            & 0.27                \\
                 &                          & Spatial-Temporal   & 0.32                \\ \hline
M$\Delta$-ValNet & Mood, $\Delta$, Valence  & Spatial            & 0.38                \\
                 &                          & Temporal           & 0.52                \\
                 &                          & Channel            & 0.50                 \\
                 &                          & Spatial-Temporal   & 0.54                \\ \hline
\end{tabular}}
\end{table*}
}

{\renewcommand{\arraystretch}{1.27}%
\begin{table*}[t]
\caption{Ablation study results with varying number of frames in the input sample.}
\centering
\label{tab:ablation_frames}
\resizebox{0.82\textwidth}{!}{%
\begin{tabular}{|c|c|c|c|c|}
\hline
\textbf{\begin{tabular}[c]{@{}c@{}}Number of\\  frames\end{tabular}} & \textbf{Model}               & \textbf{Training labels} & \textbf{Attention} & \textbf{CCC (EMMA)} \\ \hline \hline
3                                                                    & M$\Delta$-ValNet + Attention & Mood, $\Delta$, Valence  & Spatial-Channel    & 0.51                \\
7                                                                    & M$\Delta$-ValNet + Attention & Mood, $\Delta$, Valence  & Spatial-Channel    & 0.33                \\ \hline
\end{tabular}}\vspace{-2mm}
\end{table*}
}

\subsubsection{\textbf{Arrangement of the Attention Modules}}
Table~\ref{tab:ablation_placement} presents the results obtained when the spatial-channel attention module is placed inside or outside the ResNet block. The original input samples are fed to the sequential spatial-channel attention module (see Fig.~\ref{fig:attention_modules}), and the attention maps are passed as input to the mood branch in M-ValNet. For M$\Delta$-ValNet, the attention maps are passed as input to both the mood and $\Delta$ branches. Placing the attention modules inside the residual blocks enables a focus on the relevant parts of the subsequent output of each residual block. On the contrary, attention is employed only once when placed outside ResNet18, implying the focused input is passed only once in the latter case. An evident reduction in CCC is observed when attention modules are placed outside the residual block.  

\subsubsection{\textbf{Different Attention Modules}}
Table~\ref{tab:ablation_attention_modules} presents the results obtained with various attention modules. Since superior performance was observed when the modules were integrated inside the residual block, this ablation study is performed by placing various attention modules inside the ResNet block. Spatial, channel, and temporal attention modules are individually integrated, and a sequential arrangement of the spatial-temporal attention modules is integrated within ResNet18. Temporal attention in neural networks enables models to selectively focus on specific time steps, giving more importance to certain parts of the sequence. In the sequential arrangement of spatio-temporal attention, the original input is passed to the spatial module first, and the refined input is fed to the temporal one. With repsect to M-ValNet, the best CCC is achieved with temporal attention, while the lowest CCC is observed with channel attention. With M$\Delta$-ValNet, the best performance is observed for the spatial-temporal attention module, while the lowest CCC is achieved with spatial attention. These results convey the importance of temporal information for valence prediction.

\subsubsection{\textbf{Number of Frames in the Input Clip}}
Table~\ref{tab:ablation_frames} presents the results obtained for varying number of frames sampled to synthesize the input clip. Subsampling frames from a video to form the input clip is a common step used to reduce computational burden while preserving essential information for subsequent analysis. We observe that uniformly sampling seven frames reduces emotion inference performance, as compared to sampling three frames to form the clip. The comparison shows our approach of using five frames in the input clip yields the highest CCC.

%Overall, results of the ablation study confirm that using mood as a context with spatial-channel attention inside the model enhances emotion inference performance. 

%%%%%%%%%%%%%%%%%%%%%%%%%%%%%%%%%%%%%%%%%%%%%%%%%%%%%%%%%%%%%%%%%%%%%%%%%%%%%%%%%%%%%%%%%%%%%%%%%%%%%%%%%%%%%%%%%%

\section{CONCLUSION}
\label{sec::conclusion}

%Examining emotions involves not just recognising the facial expressions or vocal tones, but rather comprehending the context in which an emotion is expressed. While psychologists have theorised the interactions between mood and emotion, the computational studies on emotion inference focus only on contexts such as background, scene, place, etc. 

Different from prior studies which use background scene, place, social agents, \textit{etc.}, as contextual cues, this study explores using long-term affect (\emph{mood}) and the emotion-change information, or the valence differential ($\Delta$), as context for inferring short-term (\emph{emotional valence}). To this end, we employ the M-ValNet and the M$\Delta$-ValNet models. We use clips from the EMMA and AffWild2 datasets to predict frame-level emotion. Additionally, we integrate sequential spatial-channel attention modules with the models. Experimental results indicate that mood and emotion-change information as contextual cues are beneficial for emotion inference, and the performance is further enhanced by incorporating attention modules. 

There is mounting evidence that mood and emotion are interconnected, and influence the elicitation of each other. This study highlights the importance of considering long-term affect as context for computationally inferring short-term affect. Besides developing robust affect estimation systems, this approach has a broader impact on mental health monitoring. The use of longer sequences for encoding mood, and fusing information from multiple modalities such as audio, head position, \textit{etc.}, will be explored as part of future work.

%%%%%%%%%%%%%%%%%%%%%%%%%%%%%%%%%%%%%%%%%%%%%%%%%%%%%%%%%%%%%%%%%%%%%%%%%%%%%%%%%%%%%%%%%%%%%%%%%%%%%%%%%%%%%%%%%%
%\addtolength{\textheight}{-3cm}   % This command serves to balance the column lengths

%%%%%%%%%%%%%%%%%%%%%%%%%%%%%%%%%%%%%%%%%%%%%%%%%%%%%%%%%%%%%%%%%%%%%%%%%%%%%%%%%%%%%%%%%%%%%%%%%%%%%%%%%%%%%%%%%

{\small
\bibliographystyle{ieee}
\bibliography{references}
}

\end{document}